\documentclass[final,3p,times,twocolumn]{elsarticle}
\pdfoutput=1

\usepackage{amssymb}
\usepackage{bera}
\usepackage{listings}
\usepackage{xcolor}
\usepackage{subfig}
\usepackage{caption}
\usepackage{changepage}
\usepackage{graphicx}
\usepackage{subfig}
\usepackage{float}
\usepackage{verbatim} 
\usepackage{picins}

\definecolor{eclipseStrings}{RGB}{42,0.0,255}
\definecolor{eclipseKeywords}{RGB}{127,0,85}
\colorlet{numb}{magenta!60!black}

\lstdefinelanguage{json}{
    basicstyle=\normalfont\ttfamily,
    commentstyle=\color{eclipseStrings}, 
    stringstyle=\color{eclipseKeywords}, 
    numbers=none,
    numberstyle=\scriptsize,
    stepnumber=1,
    numbersep=8pt,
    showstringspaces=false,
    breaklines=true,
    frame=lines,
    backgroundcolor=\color{white}, 
    string=[s]{"}{"},
    comment=[l]{:\ "},
    morecomment=[l]{:"},
    literate=
        *{0}{{{\color{numb}0}}}{1}
         {1}{{{\color{numb}1}}}{1}
         {2}{{{\color{numb}2}}}{1}
         {3}{{{\color{numb}3}}}{1}
         {4}{{{\color{numb}4}}}{1}
         {5}{{{\color{numb}5}}}{1}
         {6}{{{\color{numb}6}}}{1}
         {7}{{{\color{numb}7}}}{1}
         {8}{{{\color{numb}8}}}{1}
         {9}{{{\color{numb}9}}}{1}
}

\journal{Arxiv}

\begin{document}

\begin{frontmatter}



\title{OpenTwins: An open-source framework for the design, development and integration of effective 3D-IoT-AI-powered digital twins}


\author{Julia Robles}

\author{Cristian Mart\'in}
\ead{cristian@uma.es}

\author{Manuel D\'iaz}
\ead{mdiaz@uma.es}

\address{ITIS Software, University of Malaga, Arquitecto Francisco Pe\~{n}alosa, 18, 29071 M\'alaga, Spain}

\begin{abstract}
Although digital twins have recently emerged as a clear alternative for reliable asset representations, most of the solutions and tools available for the development of digital twins are tailored to specific environments. Furthermore, achieving reliable digital twins often requires the orchestration of technologies and paradigms such as machine learning, the Internet of Things, and 3D visualization, which are rarely seamlessly aligned.  In this paper, we present a generic framework for the development of effective digital twins combining some of the aforementioned areas. In this open framework, digital twins can be easily developed and orchestrated with 3D connected visualizations, IoT data streams, and real-time machine-learning predictions.  To demonstrate the feasibility of the framework, a use case in the Petrochemical Industry 4.0 has been developed. 
\end{abstract}




\end{frontmatter}


\section{Introduction}

Industry 4.0 has revolutionized production and technological capabilities in a wide range of sectors such as manufacturing, inspection, and automotive. In modern times, the assets surrounding us (e.g., Vehicle to Everything-V2X) are increasingly connected, sharing information with each other and with the environment to optimize their performance, prevent dangerous situations, and improve safety. We can highlight two main emerging paradigms that have gone hand in hand with the industrial revolution: the Internet of Things (IoT) \cite{diaz2016state}, which has enabled the real-time monitoring and actuation of multiple physical phenomena, and artificial intelligence (AI) and machine learning (ML)\cite{de2020toward}, which have paved the way to the modelling of the behaviour of systems and processes (in some cases unknown) through data-driven approaches. It is therefore not paradoxical that, at the same time as the surrounding sources of information have increased (IoT), analytical techniques have also evolved to improve knowledge of the environment (AI/ML). As a result of this intersection, current deep learning techniques, which use a large amount of (IoT) data to model feature extraction and classification and/or detection of complex patterns in a single pipeline, can be highlighted.

The requirements for continuous analysis and prediction of the real-time behaviour of an asset and its possible future state have led to the emergence of the so-called digital twins \cite{tao2018digital}. A digital twin can be defined as a digital accurate and trustworthy representation of a physical asset \cite{rasheed2020digital} provided through continuous monitoring, prediction, and optimization for decision-making. For instance, consider a train wheel bearing, which tends to have high maintenance and production costs \cite{marquez2020designing}. A digital twin that predicts when bearings need to be repaired and/or replaced, as well as their estimated service life, could highly optimize operating costs and allow for better planning by railway companies.  Although digital twins share similarities with cyber-physical systems, such as the integration of assets into the digital world, they go a step further \cite{nazarenko2020role}, providing accurate replicas of assets that behave as they would behave in the physical world. Digital twins exploit the paradigms that have driven the Industry 4.0 to monitor the environment (IoT) and seek the continuous optimization/prediction (AI/ML) on physical assets. In many processes, in addition to data-driven models, physically‐based models provide accurate representations of asset behaviours through mathematical equations. However, these models are often difficult to obtain, so they are complemented by data-driven models such as AI/ML. Digital twins are, therefore, a combination of technologies, models, and paradigms integrated into a common interface for asset control and monitoring.  In this context, the visualization (preferably in 3D) of assets also plays a key role, as one of the main functionalities of having reliable digital twins is the simulation of the assets in unknown and extreme situations in order to evaluate their behaviour.

This orchestration of technologies and paradigms presented by digital twins requires frameworks for their continuous development and integration. Over the last few years, a large number of frameworks have emerged that enable the development of digital twins. However, in the vast majority of cases the digital twins frameworks were bounded to the domain applied \cite{conde2021modeling}. A notable open-source example is Eclipse Ditto\footnote{https://www.eclipse.org/ditto/}, one of the most widely used solutions for multi-domain digital twins. Whereas Eclipse Ditto allows for virtual abstraction of asset communications, as well as fine-grained access control management, substantial extra integration efforts are required to achieve effective digital twins. Of course, we are referring to seamless harmonization with AI/ML techniques, integration with 3D rendering engines, sensor failure detection, and integrated visualization of the ecosystem, among others. Moreover, digital twins often address individual assets, but these may be part of an global system (e.g., the bearings and the train), where orchestration of digital twins, or digital twin composition may be necessary. In this paper, we present an open-source framework based on Eclipse Ditto for the continuous development and integration of effective digital twins. This framework provides an integrated solution for the main technologies and paradigms discussed in the context of digital twins (namely IoT, AI/ML, 3D visualization), providing a unified interface for their monitoring, management, and continuous optimization. The framework has been designed on a scalable platform that allows for fault tolerance and high availability and opens the door to future extensions. The main contributions of this article are as follows:

\begin{enumerate}
    \item An integrated ecosystem based on Eclipse Ditto for the abstraction and continuous monitoring of physical assets.
    \item Seamless orchestration with AI/ML techniques and data streams for continuous optimization and prediction, such as sensor failure detection.
    \item 3D-rendering engine integration for the design and visualization of 3D-powered digital twins.
    \item As a result of the combination of the aforementioned goals, the overarching contribution of this paper is an integrated open-source framework and interface for the design, development, and continuous integration of effective digital twins composition.
    \item Finally, this paper presents the validation of the framework by means of a manufacturing use case in the Petrochemical Industry.
\end{enumerate}

The rest of the article is organized as follows.  In Section \ref{sec:related-work} related work is discussed. Section \ref{sec:architecture} presents the digital twin framework architecture and its components, whereas the implementation details are discussed in Section \ref{sec:implementation}. An evaluation of the framework is performed in Section \ref{sec:evaluation}. Lastly, our conclusions and future work are presented in Section \ref{sec:conclusions-future}.

\section{Related work}
\label{sec:related-work}

Research on digital twins has had a major impact in recent years. However, whereas some of the works deal with design and architectural aspects \cite{zheng2019application, cheng2020dt}, others focus only on specific sectors \cite{mo2021terra, dang2021cloud}. Therefore, application developers may struggle to find digital twin frameworks in the Industry 4.0, especially open-source tools for the development of effective 3D-IoT-AI-powered digital twins. Some of the most relevant works in this field are summarized below.

Kamath et al. \cite{kamath2020industrial} aim to bridge the gap between academia and industry by providing an architecture for the development of digital twins based on open-source components. Our architecture shares some similarities with their study, as some of the open-source technologies used, such as Kafka, InfluxDB, and Eclipse Ditto, are also adopted in this work. However, our architecture goes further, as it allows the integration not only of digital twins with real-time data monitoring but also with machine learning techniques (Kafka-ML) and 3D rendering. Finally, in our architecture, in addition to using open components, the whole digital twin platform and related components are available on GitHub\footnote{https://github.com/ertis-research/digital-twins-platform}.

Karan et al. \cite{shah2021construction} also consider the inclusion of open tools such as Eclipse Ditto and OpenPLC for the realization of a digital twin framework. They orchestrate a simulation model for modelling Computational Fluid Dynamics (CFD) of a portable temperature-controlled chamber. Even though the integration of simulation models is promising for combining real-world monitoring data assets and physical models, this framework lacks flexibility and is only adapted to CFD systems.

A modular digital twin framework is presented in \cite{rolle2021modular}. This framework enables the monitoring, assessing, and 3D visualization of assets and has been validated on a real manufacturing scenario. Some KIPs can be defined for measurement and alert generation through LabVIEW. The main limitation of this architecture is that all the elements are connected to a single OpenPLC server. As evaluated in the article, scalability can be a challenge in this architecture, and a possible solution is the replication of the framework structure for each process/machine to be monitored and modelled. Instead, our architecture is entirely based on fault-tolerant containers that are managed in real time through Kubernetes.

A digital twin development guide has been built around the FIWARE ecosystem in \cite{conde2021modeling}. FIWARE is a well-known ecosystem promoted by the European Commission for the development of next-gen applications in multiple sectors. FIWARE has the open-source components in place for data ingestion, data analysis, authentication, and data storage, among others, for the service realization. The solution describes a guide of how digital twins can be defined in FIWARE, but a framework adapted for digital twins has not been developed and validated as in our work. 

Khan et al. \cite{khan2020toward} propose a spiral digital twin framework, which aims at secure and reliable management of digital twin data through blockchain. This framework aims at providing a continuous and consistent synchronization between the digital twin and the real assets. A new blockchain solution, twinchain, is proposed to address the problems of merging digital twins with the blockchain technology. The spiral digital twin framework focuses on improving the reliability of digital twin information and does not provide a general-purpose framework for the development of effective digital twins as our work does. Although communications within our components are encrypted, we will also look into adopting blockchain to improve the reliability of the system in the future.

\section{Open-source architecture for the design and development of 3D IoT-AI-powered digital twins}
\label{sec:architecture}

The fundamental pillar of the architecture is the well-known framework for digital twins Eclipse Ditto. Eclipse Ditto offers an entity to model twins, the storage of the state of the twins and their events, an access control system, and the support of different types of connections that allow interaction with the twins and with other backends. Incoming messages to Eclipse Ditto, except those from the HTTP API, must follow the Ditto Protocol\footnote{https://www.eclipse.org/ditto/protocol-overview.html} format, or a payload mapping must be configured in the connection.
All the necessary elements in our architecture have been connected around Eclipse Ditto, both to make up for its shortcomings in terms of what a digital twin essentially requires and to extend its functionality. All these elements are open-source tools, and, although most of them are external projects, some of the services have been specifically developed to add certain functionality or to support the connection of incompatible elements.

\begin{figure*}[h]
  \includegraphics[width=\linewidth]{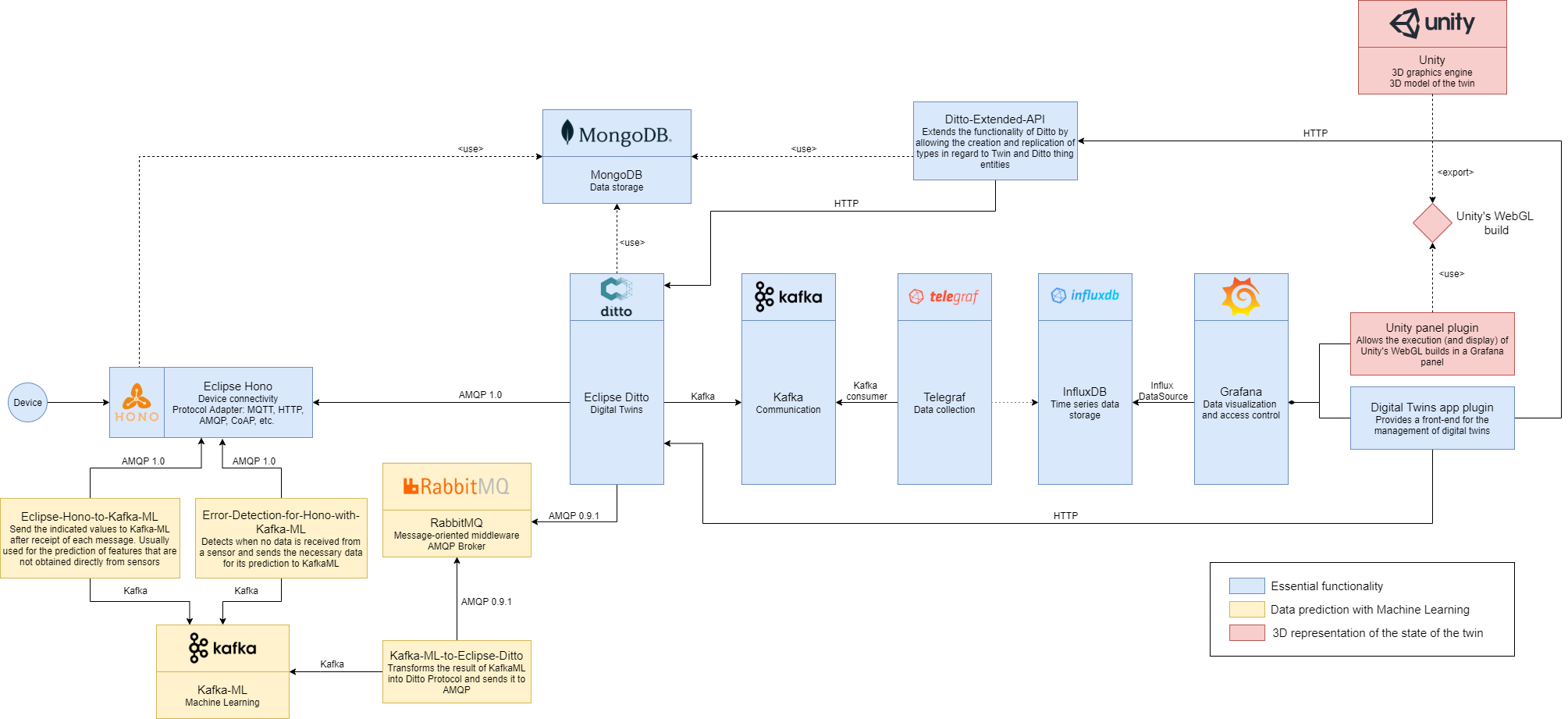}
  \caption{An overview of the 3D-IoT-AI-powered digital twin architecture. In blue the essential functions, in red the 3D representation and, in yellow the data prediction with ML.}
  \label{fig:architecture_general}
\end{figure*}

The result, of which we can see an overview in Figure \ref{fig:architecture_general}, is a microservices architecture that covers the basic needs of a digital twin platform, such as device connection, real-time data storage, and data visualization, but that really stands out for other functionalities, such as the real-time data streams prediction through machine learning and the possibility of displaying the state of the twin by means of interactive 3D representations.


The microservices architecture has a great modularity, reusability and scalability. Each of its components is independent and is responsible for a specific function. This makes it possible to replace them easily without affecting the rest of the system,  reuse some of these services in other projects, and extend the system with less difficulty.  Regarding communication between services, most of them have an API (Application Programming Interface) to facilitate interaction, although when it comes to receiving and sending real-time data, other pub-sub protocols, such as AMQP and Apache Kafka, are considered.

To ease the management of all the services in the architecture, services have been packaged in containers, and a container orchestrator coordinates them. In this case, Docker has been used as the container technology and Kubernetes as the container orchestrator. In each of the external projects, its version for Docker containers has been chosen, and all the developed services have been containerized. Moreover, encapsulating each service in an isolated environment ensures its portability and correct execution in most systems.

In Github, \footnote{https://github.com/ertis-research/digital-twins-platform} the description of the system, the installation and connection manual of the architecture, the necessary documentation for its use, and the redirections to all the services and plugins developed, (which, in addition to the code produced, also have their own installation manual and documentation of use) can be found.

The discussion of the architecture has been organized according to the main milestones, separating it into its basic functionality as a digital twin platform, the data prediction with machine learning, and the 3D representation of the state of the twin.

\subsection{Essential functionality}

The objective of this milestone was to obtain a platform in which the digital twin of any element made up of sensors can be defined. For this, it is essential the definition of the twin, to obtain and connect the IoT information, the storage of the twin data in real time series, and the possibility of consulting these data in a user-friendly way.

Although more secondary, it is also convenient to include the creation and management of twin types. This streamlines the tedious task of creating multiple twins that, while corresponding to different physical devices, have exactly the same features. For example, if we have ten identical sensors in the real world, we can define just one type and create ten twins for them.

The blue components in the architecture (Figure \ref{fig:architecture_general}) represent the part of the architecture corresponding to this essential functionality. This part is mainly composed of open-source projects but also includes two elements that have been developed to complete the desired functionality.

The main element is Eclipse Ditto, as mentioned an open-source framework for building digital twins. Eclipse Ditto does not provide any system to obtain the information sent by the devices, so Eclipse Hono\footnote{https://www.eclipse.org/hono/} will be used for this purpose. Eclipse Hono is a platform that provides several interfaces for connecting a large number of IoT devices, unifying them into a single AMQP 1.0 endpoint, where the information received can be read, and commands can be sent to trigger actions on any of the devices. It can receive information via common IoT protocols, such as MQTT, AMQP, HTTP, and CoAP, and custom adapters. This is the recommended tool to work with Eclipse Ditto, and, thanks to the Eclipse cloud2edge package\footnote{https://www.eclipse.org/packages/packages/cloud2edge/}, the integration of the two is very convenient.

On the one hand, the Eclipse Ditto Thing entity can be used in different ways, so after studying different options, such as considering a Ditto Thing as a complete twin including each sensor as a feature of it, a design decision has been made to assign a Ditto Thing to a single entity or sensor and to create parent-child hierarchies between these entities. These have been implemented in a service called \textit{Ditto-Extended-API}
, which can be considered as a layer above Eclipse Ditto and which provides an API that replaces the one offered by this technology. In addition to verifying that the specified constraints are satisfied, this service adds all the respective functionality to the twin types allowing both their management and the creation of twins from them.

On the other hand, another feature that Eclipse Ditto lacks is the storage of the twin state at different time instants. To solve this, InfluxDB\footnote{https://www.influxdata.com/products/influxdb-overview/} has been chosen as a time-series database. This is a well-known database with a large community and is ideal for processing sensor data. To collect the data, we have considered Telegraf\footnote{https://www.influxdata.com/time-series-platform/telegraf/}, a plugin-driven server that provides support for a large number of data sources and with which we can configure the data ingestion into the database. As there is no possibility for Telegraf to collect the data directly from Eclipse Ditto, because neither technology implements a broker of the protocols it supports, we need an intermediate element that allows its connection. Apache Kafka\footnote{https://kafka.apache.org/}, one of the best-known streaming and processing platforms for real-time data, is a very good alternative since Telegraf has a Kafka-consumer\footnote{https://www.influxdata.com/integration/kafka-telegraf-integration/}
plugin available, and Eclipse Ditto does have got the option to publish events in a Kafka topic.

Finally, Grafana\footnote{https://grafana.com/} has been chosen to act as the front-end, i.e., the user interface for end-users. This technology provides support for metrics visualization from the most popular databases, including InfluxDB. It allows making queries in the language defined by the chosen data source and presenting the result in different types of interactive panels. These panels are part of dashboards, which can be modified to the user's liking. It also includes an access control system through roles. Another of its strong points, and one of the most important for the project, is that it allows the creation of personalized panels and the inclusion of any type of functionality by means of plugins, providing the libraries and documentation necessary for this.

A plugin app\footnote{https://github.com/ertis-research/digital-twins-plugin-for-grafana/}, called Digital Twins, has been developed for Grafana. This plugin adds a graphical interface for managing twins and types in the same application that is already used in the front-end for querying the status of the twins, thus unifying all the functionalities of the platform. Its functionality consists, basically, in making calls to the Ditto-Extended-API service in a user-friendly way.


\subsection{Data prediction with Machine Learning}

This milestone aims to achieve the integration of the platform with machine learning algorithms. This might be useful for digital twins to predict their next state or a situation of failure as, for instance, the values that a sensor should return in case no real data are received from it, either because it has been switched off or because it has got some kind of failure.

The part of the architecture that is in charge of achieving this objective corresponds to the yellow components in Figure \ref{fig:architecture_general}. In this part, the main component is Kafka-ML \cite{martin2022kafka}, which will be in charge of machine learning life cycle management and complements this architecture. In addition, in order to integrate it with Eclipse Hono and Eclipse Ditto and fulfil the required functionality, three specific services have been developed: \textit{Eclipse-Hono-to-Kafka-ML}, \textit{Error-Detection-for-Hono-with-Kafka-ML} and \textit{Kafka-ML-to-Eclipse-Ditto}.

Kafka-ML is an open-source framework \footnote{https://github.com/ertis-research/kafka-ml/} developed by our group 
that manages the life cycle of ML/AI applications in production environments through continuous data streams. Unlike traditional frameworks that work on datasets or static files, Kafka-ML allows both training and inference with continuous data streams, enabling users to have a fine-control of the ingestion data in popular ML frameworks such as TensorFlow and PyTorch. Kafka-ML currently supports popular ML frameworks such as TensorFlow and PyTorch and through its user-friendly Web UI allows the management and deployment of ML models, from their definition to final deployment for inference.   

One of the main purposes in the area of machine learning focused on digital twins is the prediction of the future states of the twin, which may be of considerable utility if control or improvement of the element that the twin represents is sought. It also aims to predict certain features or values of the twin that cannot be measured or obtained directly or in any accurate way. As mentioned above, these models will be deployed in Kafka-ML, so for their connection with Eclipse Hono a service has been developed, which has been named \textit{Eclipse-Hono-to-Kafka-ML}\footnote{https://github.com/ertis-research/eclipse-hono-to-kafka-ml}
. This service constantly reads from the Eclipse Hono endpoints corresponding to the tenants (entity that allows the logical partitioning of devices into unique groups) that contain devices whose data must be sent to these deployed Kafka-ML models. When a message arrives from one of these devices, the service processes the data to comply with the required format and automatically sends it to the respective Kafka-ML input topic, where the model will process the data to obtain a prediction as a result. Then, the digital twin is updated with the prediction. 
Remarkably, when using the prediction of future states of the twin, the main digital twin is not updated, but a copy of it is made in Eclipse Ditto containing the predicted state. This copy will be considered the same twin but advanced a certain time.

Another of the strong points of this part is to detect when a sensor is not sending its data and to act in consequence. To this end, a new service \textit{Error-Detection-for-Hono-with-Kafka-ML}
\footnote{https://github.com/ertis-research/error-detection-for-eclipse-hono-with-kafka-ml/} 
has been created, which basically reads the information received by the specified Hono tenants and checks that the devices that are part of it are sending their data in accordance with their periodicity. In case one of them does not send the data when it is due, the service will send the last values received from that sensor plus other necessary data, such as the date, to a Kafka-ML trained model with historical data. Kafka-ML will predict the next state of the system to avoid a service interruption due to the sensor failure until the sensor is available again.

In the other direction, the data generated by Kafka-ML have to be consumed by Eclipse Ditto. Initially, the idea was to take advantage of the payload mapping functionality provided by Ditto to Kafka-ML, creating a source connection to each of the Kafka-ML output topics where models send the predictions, and mapping the information received so that it could be supported by Eclipse Ditto. 
This was not viable, since, at the time of the development of this platform, Eclipse Ditto had not implemented the connection with Apache Kafka acting as a data source. That is why it was decided to build an intermediate service \footnote{https://github.com/ertis-research/kafka-ml-to-eclipse-ditto/}
that would read the information from Kafka, map it to Eclipse Ditto Protocol, and publish it to a message broker that Eclipse Ditto could connect to, in this case RabbitMQ\footnote{https://www.rabbitmq.com/}, which use AMQP 0.9.1.


\subsection{3D representation of the state of the twin}

An important aspect of digital twin platforms is the representation of the data. In the most relevant industrial platforms, it is common to find 3D representations of the twin that make it much easier to understand its information and the component that is being consulted at any given moment. Achieving this 3D representation of the twin is the main objective of this milestone. 

The architecture that has been designed for 3D visualization corresponds to the red components in Figure \ref{fig:architecture_general} and basically consists of the creation of a panel plugin for Grafana that allows the display of a 3D model developed with Unity\footnote{https://unity.com/} with which, moreover, it will be possible to interact in both directions.

Unity is a software that mainly focuses on video game development, although it can also be used in other contexts. It is one of the most popular graphics engines with the largest community. Although it is not an open-source tool, it can be used free of charge for personal use or for low-budget projects. 
This technology allows assigning a certain behaviour to 3D objects through the use of scripts and interacting both with the user and with other elements in the environment. A wide variety of formats are available for importing 3D models, including Blender. Blender is an open-source 3D creation suite that also has a large community and will be of help when creating or modifying 3D objects. Unity also allows the project to be built in several formats, including a specific one for web rendering called Unity WebGL. Compilation in this format will also be of essential importance in this part of the architecture.

Having the 3D representation of the twin built in this format, the next step is to run it in Grafana, the tool we have chosen as the front-end of the platform. To do this, we developed a panel plugin. This type of plugin allows the creation of a custom panel that can be included in any dashboard. It usually represents or uses data series, which will depend on the query and the data source introduced by the user in the panel configuration. Additionally, it can be implemented in such a way that it provides the user with a series of custom options in its configuration by means of which its behaviour or visualization can be easily modified. 

As well as rendering a Unity WebGL building, the plugin panel must allow interaction with Grafana and with the user in both directions. In other words, the user's interaction with Grafana will be reflected in the 3D model, and, in turn, the user's interaction with the representation may have repercussions on the information displayed by other panels of the dashboard. This plugin is called Unity panel plugin.\footnote{https://github.com/ertis-research/unity-plugin-for-grafana/}. 

\section{Development and implementation}
\label{sec:implementation}

Next the complete development of the platform will be explained in detail, including the connection of the different technologies, their relationships, and the design of the developed services mentioned above.

\subsection{Linking basic architecture}

For the deployment of the different technologies in Docker containers managed by Kubernetes, we have chosen to use Helm\footnote{https://helm.sh/}. Helm is a package management tool for Kubernetes that makes it much easier to install any technology by assembling its recommended deployment with a single command, saving the manual creation of Kubernetes objects. Its packages are called charts, and can be easily customized by modifying the necessary parameters in a YAML values file.

Eclipse Ditto and Eclipse Hono have been installed using the Eclipse cloud2edge package. It installs both technologies so that they can communicate and establishes a connection between them with an example device. In order for the installation to work correctly, the Grafana and Prometheus options of Eclipse Hono had to be disabled, and it was necessary to create a Kubernetes persistent volume to correspond with the device registration service of the same tool, thus enabling the persistence of data. 

\begin{figure}
  \includegraphics[width=\linewidth]{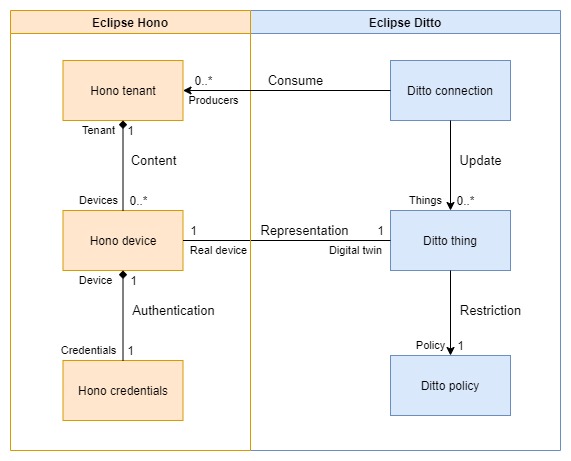}
  \caption{Relationship between Eclipse Ditto and Eclipse Hono}
  \label{fig:relationship_dittoandhono}
\end{figure}

Just with this, we could create the model of a digital twin and update it with the information provided by the sensors it represents. At this point, it is important to emphasize the relationship between the two tools, which is illustrated in Figure \ref{fig:relationship_dittoandhono}. In Eclipse Hono, devices are defined within a tenant, simply by assigning them an identifier. A tenant is just a grouping of devices. For the actual device to be able to send data to this entity, it has to use this identifier and some credentials that must be previously added. We can access the data of a device by consulting in the AMQP endpoint the address corresponding to the tenant to which it belongs. On the other hand, in Eclipse Ditto we can create Ditto things entities, which will be restricted by policies and whose identifier (thingId) will be its namespace plus an identifier, following the namespace:id format. For the Ditto thing to receive the update messages, the Ditto connection to the Eclipse Hono AMQP endpoint must be established as indicated in the cloud2edge package, and the thingId of the corresponding Ditto thing must be established as the identifier of the Eclipse Hono device.

Regarding the sensors, IoT devices are expected to send messages to Eclipse Hono in Ditto Protocol format, although in case they cannot, a mapping can always be established in the connection to Eclipse Ditto. Messages should be sent to the Eclipse Hono service in charge of the chosen connection protocol and the connection should be authenticated with specific credentials, which will indicate to Eclipse Hono the device to which they refer.

Following the architecture, the next step was to publish the status of the twin in an Apache Kafka topic. In this case, resources have been reused to install both Apache Kafka, and CMAK\footnote{https://github.com/yahoo/CMAK}, a graphical interface to manage Kafka clusters. Once ready, a topic has been created to receive the information coming from Eclipse Ditto. Eclipse Ditto allows the creation of target connections to send the twin information to external systems, including Apache Kafka brokers. When defining it, it is necessary to specify which data topics to subscribe to. In our case, it was the events of the twin, since Ditto launches one after each update and includes in the message the new state of the twin. It was also necessary to specify in the authorization context a valid user that has at least read permissions on the twin. We can consult the existing users and their permissions in the associated policy. Once the connection is created in Eclipse Ditto, every time the twin is updated, its new status will be published in the Kafka topic we created.

To install InfluxDB, the Helm version has also been chosen. NodePort has been set as service type and a specific password has been assigned to the administrator, thus avoiding future problems if the package needs to be restarted. During installation, it was also necessary to create a persistent volume related to a custom storage class. Although InfluxDB provides a very intuitive interface, which is the one that will be mainly used, it is interesting to install and configure also the Influx CLI for certain exclusive functionalities such as user management.

Once InfluxDB was configured, we had to collect the data exposed in the Kafka topic created earlier. For this purpose, Telegraf is used, a tool that was also installed using Helm. To indicate the selected plugins and their configuration to Telegraf, it is required the use of a custom file of values in YAML format during the installation. In this file, in addition to the parameters corresponding to the deployment in Kubernetes, there is a configuration section. At this point, it is essential to review the tool's documentation to see what parameters are required by each of the plugins, regardless of whether they are input or output. In our case, as output we have InfluxDB in its second version, which in one of its fields will require a token with write permissions previously created in InfluxDB. As input, we have a Kafka consumer, to which we specify the address of the broker and the topic to which it should subscribe. After applying this configuration, our InfluxDB instance should already be receiving the metrics from Kafka.

Finally, there is the connection between InfluxDB and Grafana, which will allow queries to be made to an InfluxDB database from Grafana panels, which will display the result obtained. First, the Grafana chart has been installed, activating persistence. 
Once the tool is ready, in the same interface, InfluxDB has been configured as the data source, indicating that it is the version that uses Flux as the query language and filling in the rest of the requested data. Once the tool approves the connection, we can now represent the twin information in Grafana, thus finishing the connection of all the existing technologies and moving on to the development of the extra functionalities.

\subsection{Development of Ditto-Extended-API}

Eclipse Ditto offers us the Ditto Thing entity, which always belongs to a namespace and is basically composed of an identifier and a series of attributes and features. The attributes correspond to the static part of the entity, whereas the features correspond to the dynamic part. In this way, this technology gives us full freedom to define how we want our models to be and how we want to use the tool. If a Ditto thing corresponds to a single sensor, we can store manufacturing information, for example, as attributes, whereas the data it receives will be perceived as features. Another option is that the Ditto Thing entity encapsulates a set of sensors, separating the data of each one within its features.

\begin{figure}
    \centering
  \includegraphics[width=0.9\linewidth]{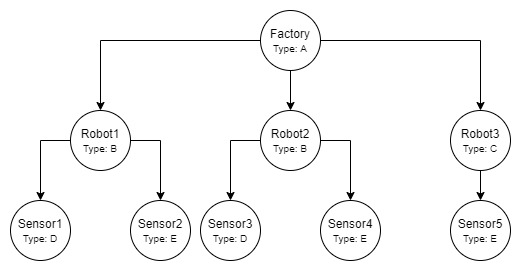}
  \caption{Example of digital twin represented as a tree.}
  \label{fig:tree_twin}
\end{figure}

In order to understand the design decision taken, it is interesting to first clarify certain concepts about digital twins. A digital twin can be considered both an entity that receives information from a single device, and one that is composed of other entities, which can also be understood as twins. A twin could then be represented as a tree, where each leaf is the representation of a single sensor. Thus, a factory that has three robots that each have particular sensors could be contemplated as shown in Figure \ref{fig:tree_twin}.

%

\begin{figure}
  \centering
  \includegraphics[width=0.6\linewidth]{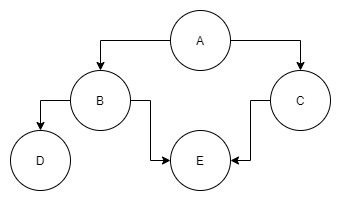}
  \caption{Example of types of twins represented in graph form.}
  \label{fig:tree_type}
\end{figure}

As for the twin types, in the example of the factory, if robots 1 and 2 are the same model, we could create a type to facilitate their creation in case more robots of the same type are added to the factory. In addition, the sensors that compose them should also form their own types since, for example, sensor 2 (and therefore sensor 4) can be of the same model as sensor 5, even if they belong to different robot types. In this way, the twin types constitute a directed graph without cycles, which in the case of the example could be represented as in Figure \ref{fig:tree_type}.


\begin{figure}
  \includegraphics[width=\linewidth]{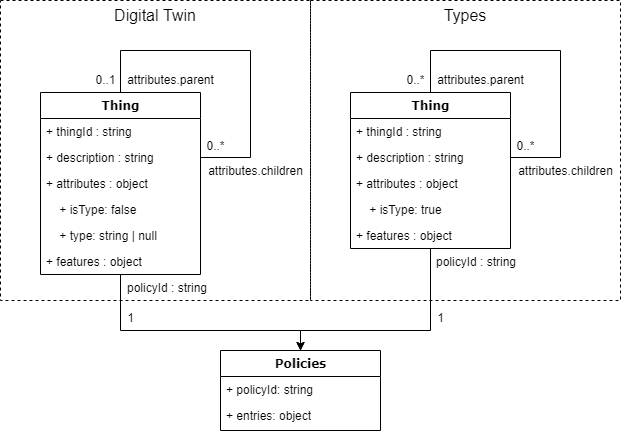}
  \caption{Use of Ditto entities by ditto-extended-api.}
  \label{fig:entities_api}
\end{figure}

Following these guidelines, a small service has been designed that uses the Ditto Thing entity in a way that complies with these properties, acting Eclipse Ditto as its database. Figure \ref{fig:entities_api} shows a diagram of how this application makes use of the Ditto entities. Basically, we have taken advantage of the fact that the attributes section of the Ditto Thing corresponds to any JSON object to include, apart from the desired static information, a series of specific attributes. Foremost, a boolean attribute \textit{isType} has been added to indicate whether the thing corresponds to a type. Depending on this value, which in case the attribute does not exist will be taken as false, there will be some or other restrictions. In case the thing does not correspond to a type, you can have a text attribute \textit{type} indicating the identifier of the type that the twin is, if any. On the other hand, a system of parents and children has also been established. In both cases, things can have any number of children, but this is not the same for parents: twins can have at most one parent, whereas types can have more than one. This is because twins correspond to real instances that, logically, can only be in one place, whereas types correspond to abstract instances that can be part of several distinct instances. To satisfy this, for twins the \textit{parent} attribute corresponds to a text field containing the identifier of the parent, whereas for types this attribute is a JSON object where each key is the identifier of one of its parents and has a value of 1. The \textit{children} attribute, which is shared by both entities, has a similar format to the latter, with the difference that the value of each child identifier of a twin must be strictly 1 and that of types can be equal to or greater than 1. These four attributes cannot be added or modified manually, and an error message will be received when attempting to do so.


\begin{figure*}[t]
  \includegraphics[width=\linewidth]{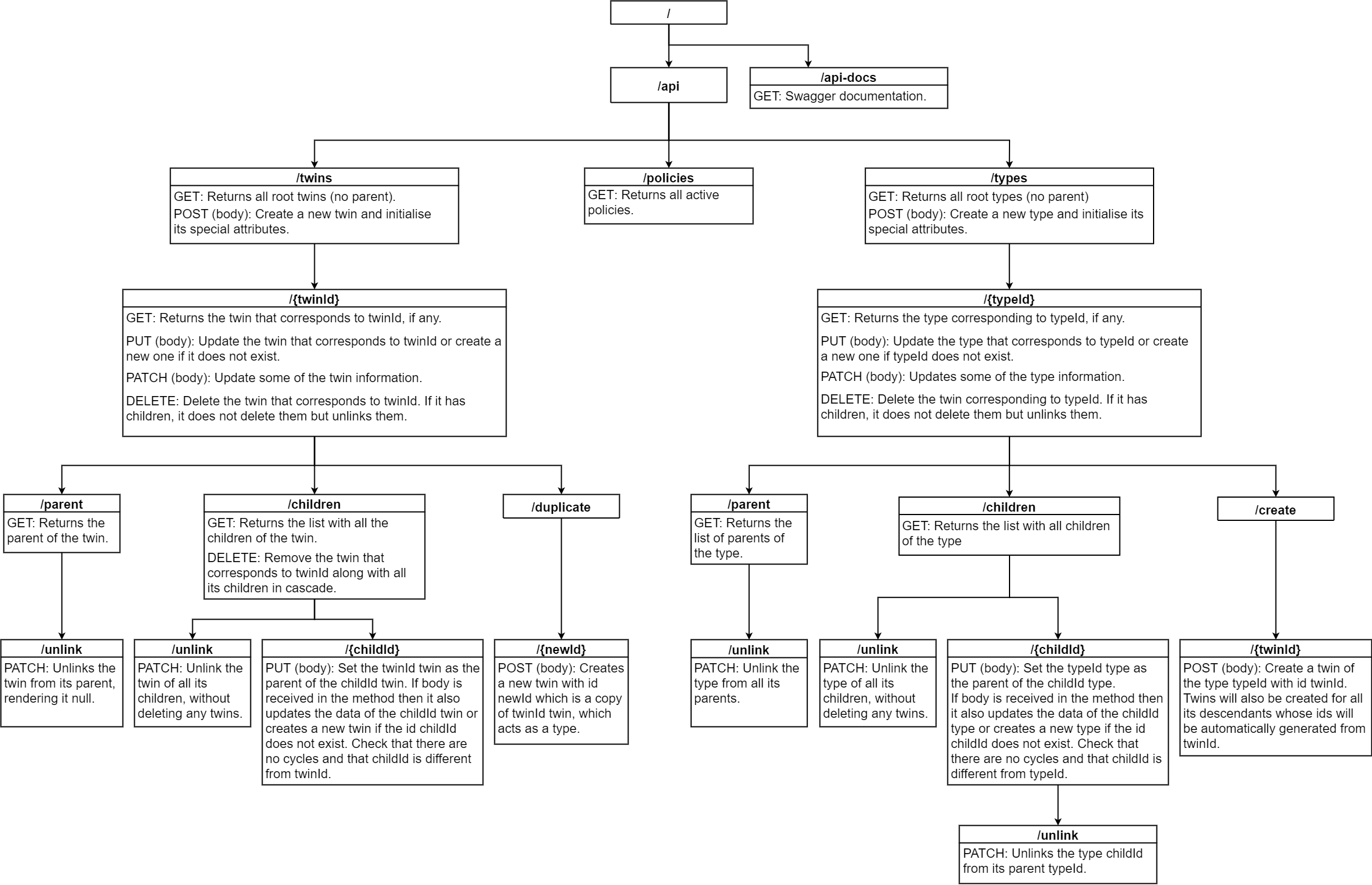}
  \caption{Available API calls in ditto-extended-api.}
  \label{fig:ditto_extended_api}
\end{figure*}

The service acts as a layer on top of Eclipse Ditto, since it provides an API that is very similar to Eclipse Ditto but takes all these restrictions in consideration. So, if we want to create a new thing the service will automatically initialize these specific attributes depending on whether we want to create a single twin, a type, a twin from a type, or a twin that is a child of another twin. When editing, it is always taken into account that the specific attributes are not modified and the parent-child relationship can only be modified by means of specific calls. On the other hand, the things returned by the queries will hide the isType attribute, as these will be considered as separate entities. Regarding the deletion of a twin, it will be possible to choose between deleting only that entity and leaving all its children without a parent or deleting that entity and all its children in cascade. In the case of types, only the first option can be selected. The interaction with the Ditto policy entity remains as it is, although a query is added that returns a list of all existing policies. Figure \ref{fig:ditto_extended_api} shows an overview of the calls currently provided by the API.

The service has been developed with the NodeJS framework, whose programming language is JavaScript. This open source framework works at runtime and provides great performance in the development of server-side tools. 


After the implementation and testing, the service has been containerized for Docker and deployed on Kubernetes. It should be noted that ES6 has not been used for the implementation because it was incompatible with the version of Kubernetes used in our cluster. Finally, to use the service, the URI where Eclipse Ditto is deployed, valid credentials to interact with it and the MongoDB URI for the Ditto policy database must be set as environment variables. 



\subsection{Digital Twins app plugin for Grafana}

Grafana offers three types of plugins: data source plugins, panel plugins and application plugins. In this case, the latter option is of interest to us, as it will allow us to add a new section in the left bar of the Grafana application, in which we will have a page with as many tabs as necessary. Within each one of them it is possible to add any type of functionality, although it must be taken into account that Grafana has certain defined styles, so it is recommended to use them to maintain consistency.
In addition to this page, every plugin includes a configuration section, which is also a tabbed page, where we can typically find the plugin readme, the option to enable/disable it and, if there are any, all the configuration options necessary for the plugin to work.

This plugin should act as a front-end to the platform, providing the user with an interface that brings together the different functionalities of the platform. Specifically, it must allow the management of digital twins and types through the Ditto-Extended-API service, the administration of policies and connections directly in Eclipse Ditto, the connection with Kafka-ML through the two services developed and the creation of devices in Eclipse Hono. All these actions will be done through API calls, in order for the plugin to have as minimal business logic as possible.

To accomplish this, a tab on the main page will be assigned to each of the entities: twins, types, policies, and connections. This tab will modify its appearance depending on the parameters received in the URL. Initially it will display a list of all the elements of the entity, although, in the case of twins and types, it may be restricted to elements that do not belong to any other (no parents) as they are considered the main entities. All elements may be selected for query, edit or delete. When querying a twin or type, the information of the element will be displayed, as well as lists that allow querying its parents and children. When consulting a twin, it will also be possible to manage its relations with Eclipse Hono and Kafka-ML. For these relations, a valid connection with each technology must be selected. In the connections section you can create Eclipse Hono tenants and connect them, or others already created, to Eclipse Ditto.

The configuration section will be filled in only with fields to enter the addresses and credentials necessary to use the different technologies that make up the platform. If these fields are not filled in, the application will not be able to function.

In terms of implementation, Grafana provides a template for each type of plugin, which can serve both as a basis and to give an idea of how the plugin works. All plugins have a series of essential files so that Grafana can detect the plugin and configure it. ReactJS, an open source, component-based JavaScript framework, is used for coding. This framework's main purpose is to create user interfaces for single page applications. Each tab can be assigned a ReactJS component. In our case this component will be used as a component selector to be able to have the different modes of an entity: consulting its elements, consulting an element and editing an element. This selection will be made by means of a parameter included in the URL. On the other hand, Grafana provides a series of libraries that will allow a better integration with the tool. The use of the grafana/ui library is essential to maintain the consistency of the forms and other visual elements of the plugin with respect to Grafana.

Once the plugin has been built correctly, its code must be added to the Grafana plugins folder and activated in the corresponding section of the application. As soon as the required fields in the configuration have been filled in, the plugin will be ready for use.

\subsection{Error detection for Eclipse Hono with Kafka-ML}

Eclipse Hono allows the collection of data received from several devices in a single AMQP 1.0 output, in which they can be consulted in the endpoint corresponding to the tenant to which the device belongs. This endpoint has the form \textit{telemetry/tenantid}, where \textit{tenantid} is the tenant identifier. The problem with Eclipse Hono is that its functionality does not include any way to control that certain devices are sending data periodically. Its closest feature is the \textit{ttd} field that is received as a device notification. This allows you to indicate whether a device is available indefinitely, unavailable, or to specify how many seconds it will be available after receiving its last message. This is useful for checking that the device is available before sending a message to it, but it is not really relevant for our purposes as it cannot be manually configured for the AMQP and MQTT protocols and, even if it could, the frequency at which each device will send its data is unknown.

The functionality sought is just that, the constant verification that the device is sending data periodically. Furthermore, if no data is received in the usual time interval, it will be assumed that the sensor has some kind of error and predicted data will be produced by Machine Learning to cover the lack of information. This error detection cannot depend on the reception of any special message from the sensor, as the service must be susceptible to all types of errors, including those that may cause a lack of connection.

\begin{figure}[h]
  \includegraphics[width=\linewidth]{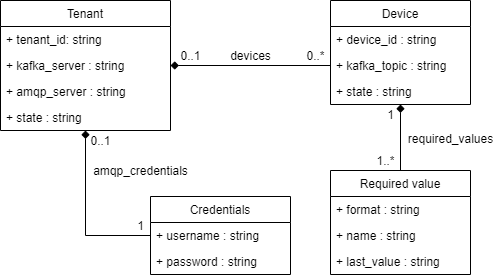}
  \caption{Entities of Error detection for Eclipse Hono with Kafka-ML}
  \label{fig:error_detection_bd}
\end{figure}

For this purpose, a service has been designed that will launch a thread for each Eclipse Hono tenant to be monitored. Within these tenants, the devices indicated by the user will be supervised. In case no data is received from any of them, the last values obtained will be sent to Kafka-ML as well as the information required for the prediction, following the format specified by the user. The control of tenants and devices can be easily deactivated and activated, avoiding the elimination of the rest of the data. A MongoDB database will be used to store all the information needed to establish the connections and produce the Kafka-ML entries. The exact content of the information to be provided for each tenant is shown in Figure \ref{fig:error_detection_bd}.

\begin{figure}[h]
  \includegraphics[width=\linewidth]{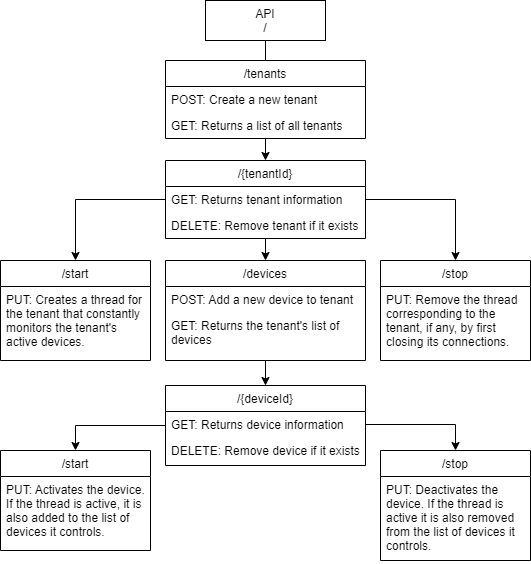}
  \caption{Endpoints of Error detection for Eclipse Hono with Kafka-ML.}
  \label{fig:error_detection_api}
\end{figure}

The management of the threads will be done through an API, which allows creating, querying, deleting, activating and deactivating tenants and devices. The available calls can be found in Figure \ref{fig:error_detection_api}. Once the application is started, all the threads corresponding to the tenants marked as active will be launched, and in the event that the application is closed, all threads and connections will be eliminated beforehand.

In each of the threads, two connections are established: one with the Eclipse Hono endpoint corresponding to the tenant and another with the Kafka-ML tool through a Kafka connection. In addition, a perpetual timer, initially off and without a defined interval, is assigned to each of the active devices belonging to that tenant. This timer has a function associated with it that must be triggered once the time has elapsed, and that will be responsible for sending the corresponding message to Kafka-ML. Once the thread is initialized, it will enter a loop until it receives an event indicating that it must stop and close the connections. The thread will act as a receiver of the AMQP endpoint, constantly waiting for messages to be received from it. Once it receives a message, it will check if it comes from one of the active devices and, if so, it will be processed. If not, the message is simply ignored. The message processing consists of saving the received values in their respective last\_value fields in the required\_values array of the device and restarting the corresponding timer with the time interval between this new message and the last one received by this same device, rounding it up and adding 0.2 seconds for uncertainty. If this is the first message received from the device, it shall be started instead of reset. The time begins to count from that moment and there are two options: a new message is received from the device that restarts the timer before it jumps, or the time runs out, the corresponding message is sent to Kafka-ML and the timer is restarted with the last set interval. In the latter case, the timer shall keep restarting and sending messages to Kafka-ML until a new message is received from the device. The interval between this message and the previous one shall not be taken into account as the difference may be disproportionate and inconsistent, so the last set interval shall be maintained.

When sending the Kafka-ML message, the required\_values array of the device is consulted and an array is created with the specified elements in the indicated format and sorted in order of appearance. To indicate a time value, the name of the value must be defined as the name of the field in the datetime library preceded by the symbol \$. For example, to indicate the year, the value name must be \$year. For the rest of the values, the last values received will be used, which must be stored in the same object. After construction, this array will be sent as bits to the Kafka-ML tool. In this way, the values required for, for example, a sensor that has to send an array in float64 format with the format [year, month, day, temperature, humidity] would correspond to the object in Figure \ref{fig:requiredvalueexample}.

\begin{figure}
\begin{lstlisting}[language=json,firstnumber=1]
"required_values": [
  {
    "format": "float64",
    "name": "$year"
  },
  {
    "format": "float64",
    "name": "$month"
  },
  {
    "format": "float64",
    "name": "$day"
  },
  {
    "format": "float64",
    "name": "temperature",
    "last_value": null
  },
  {
    "format": "float64",
    "name": "humidity",
    "last_value": null
  }
]
\end{lstlisting}
\caption{Example of the required\_values field}
\label{fig:requiredvalueexample}
\end{figure}

This service has been implemented using the Flask framework, whose programming language is Python, as it provides great facilities when creating APIs and, unlike NodeJS, allows multithreading. The main library is the Python threading library, which provides threads and timers. Of the remaining libraries used, the most relevant are those that allow the different connections: kafka-python for the Kafka producer, python-qpid-proton for the AMQP 1.0 receiver and Flask-PyMongo for working with the MongoDB database. Also, remarkable is the Marshmallow library that checks that the data sent by the user complies with the specified schema before inserting it into the database. Once the implementation of the service is finished and after passing the relevant tests, it has been containerized for Docker and deployed in Kubernetes.

\subsection{Eclipse Hono to Kafka-ML}
The aim of the Eclipse Hono to Kafka-ML service is to automate the input of sensor data for other Machine Learning models deployed in Kafka-ML, such as those capable of predicting future states of the twin or features of the twin that cannot be measured.

The design and implementation of this service are practically identical to the Error detection for Eclipse Hono with Kafka-ML service described in the previous section, so its explanation will be omitted. The design and construction of the API, the management of the threads and connections and the mapping of the message received from Eclipse Hono to Kafka-ML are the same as described above. This could be considered a simplification of the previous service, as the difference lies in that it sends a message to the Kafka-ML input topic after each message received from an active device, regardless of the time interval between them. For this reason, the \textit{ditto\_message} field does not store the last received values of the properties, as all messages received from the indicated devices are mapped and sent directly to the corresponding Kafka-ML input topic.

\subsection{Kafka-ML to Eclipse Ditto}

Eclipse Ditto supports receiving messages through various types of connections. At the time of the development of this platform, the connection with Kafka was under development, leaving available the creation of connections with AMQP, MQTT and HTTP. Concerning the messages, Eclipse Ditto requires them to be in Ditto Protocol format in order to understand what action to execute and on which twin to perform it. In case the format in which the connection producer sends messages cannot be changed, a payload mapper can be applied to the connection. Eclipse Ditto integrates several types of mappers, although the most flexible is the JavaScript mapper. On the other hand, Kafka-ML publishes in the output topic the predicted values in array format and currently does not provide any way to set a specific format.

In our case, it is necessary to receive the data predicted by Kafka-ML so that the state of the corresponding twin is completely or partially updated. Since the direct connection of Eclipse Ditto to the Kafka-ML output topic is not possible, a service is needed that collects the information from that topic, transforms it to Ditto Protocol following a given format and sends it through one of the connections provided by Eclipse Ditto. This service can be replaced in the future by direct connections to the Kafka-ML output topics that include their respective JavaScript payload mapper.

\begin{figure}
  \includegraphics[width=\linewidth]{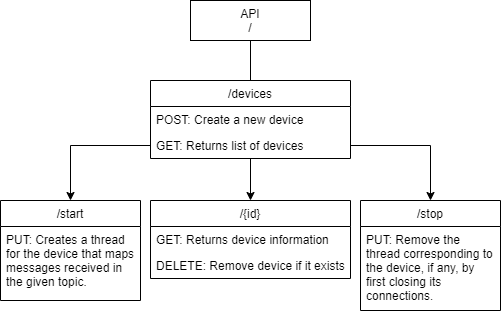}
  \caption{Endpoints of Kafka-ML to Eclipse Ditto.}
  \label{fig:kafkamltoditto_api}
\end{figure}

\begin{figure}
  \includegraphics[width=\linewidth]{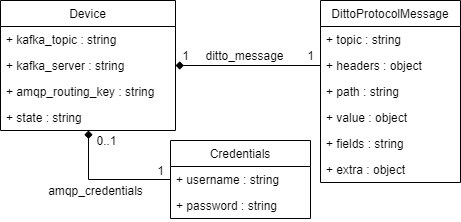}
  \caption{Entities of Kafka-ML to Eclipse Ditto.}
  \label{fig:kafkamltoditto_bd}
\end{figure}

The service, called Kafka-ML-to-Eclipse-Ditto, has a similar design to the one explained in the previous section. In general, it consists of creating a thread for each Kafka-ML output topic from which we want to extract information, transforming it and sending the result to an AMQP 0.9.1 broker. These threads can be managed through an API, which, as shown in Figure \ref{fig:kafkamltoditto_api}, allows them to be created, queried, deleted, activated and deactivated. The IP of the AMQP broker will be the same for all threads and will be set in the service environment variables. Each thread requires for its operation the information indicated in Figure \ref{fig:kafkamltoditto_bd}, which consists of the necessary data for the connection with Kafka and AMQP, the initial state of the thread and the schema of the message that will be sent to Eclipse Ditto. This schema will be a JSON object in Ditto Protocol format, in which the user has to mark in the value section the place where the values received from Kafka-ML should go. To do so, the number corresponding to the position of that value in the array received from Kafka-ML will be indicated between braces. For example, if we have a digital twin of a temperature and humidity sensor and Kafka-ML sends us an array whose first value is the predicted temperature and its second value is the predicted humidity, the ditto\_message field would be as shown in Figure \ref{fig:dittomessageexample}.

\begin{figure}
\begin{lstlisting}[language=json,firstnumber=1]
"ditto_message" : {
  "topic": "test/DHT22/things/twin/commands/modify",
  "path": "/features",
  "value": {
    "temperature": {
      "properties": {
        "value": "{0}"
      }
    },
    "humidity": {
      "properties": {
        "value": "{1}"
      }
    }
  }
}
\end{lstlisting}
\caption{Example of the ditto\_message field}
\label{fig:dittomessageexample}
\end{figure}

As for the threads, each of them creates a Kafka consumer for the topic to which it corresponds and an AMQP connection to a RabbitMQ broker. Once initialized, the thread will constantly check the topic for new messages until it receives an event that forces it to stop and close the connections. After receiving a message, the elements mentioned in a copy of the given schema will be replaced by the received values and the resulting message will be sent via AMQP to the queue indicated by the user.


As with the design, the implementation is quite similar to the previous service. Flask, Flask-pymongo, threading and Mashmallow are also used for the same purposes as those described in the implementation of the Error-detection-for-Hono-with-Kafka-ML service. On the other hand, the kafka-python library is used in this case to create a Kafka consumer capable of reading the Kafka-ML output topic and the pika library is used to establish the AMQP 0.9.1 connection. It should be noted, on the other hand, that for the deployment of the RabbitMQ instance in Kubernetes, its version of Helm was used, being necessary to activate persistence, assign a persistent volume claim previously created, activate the volume permissions and assign the type of service as NodePort.

Once implemented, tested, containerized and deployed, the service can be used. In order for Eclipse Ditto to be able to receive the generated messages, an AMQP 0.9.1 connection has been established between it and RabbitMQ, filling the source address part of the connection with the name of the AMQP queues to be consumed. In addition, a new subject has been added to the policy that fulfils the twin whose data we predict, and this subject has been included in the authorization context of the connection. To create the necessary credentials for the connections in RabbitMQ, the interface provided by the tool itself has been used. Moreover, in order to differentiate the values coming from Eclipse Hono from those coming from Kafka-ML, the Telegraf configuration has been updated to include the ditto-originator field of the message header as a tag key. After configuring this, creating the necessary entity and activating the thread, the updates for the twin should be received without any problem.

\subsection{Unity panel plugin for Grafana}

As previously explained, Grafana acts as the front-end of the platform. Its main functionality is to create dashboards composed of panels that will represent in a certain way the result of a query to a data source. Most of the default dashboards available in the tool focus on the representation of the data by employing some kind of graph. For the creation of custom dashboards, Grafana offers support for the development of panel plugins. This type of plugin is similar to the app plugin developed in a previous section, with the difference that it offers the data resulting from the query entered, and the panel options selected by the user to build the representation.

The goal here is the 3D representation of the digital twin and its state. It is also desired to be able to interact with the representation in such a way that this interaction influences the data displayed on the Grafana dashboard and, likewise, that the representation is affected depending on the state of the twin. At the moment, Grafana does not have any panel plugin that allows this functionality. Therefore, to satisfy this requirement, a plugin panel will be created to display a 3D model developed in Unity and interact with it. Although its development will focus on its use in the area of digital twins, the aim is to abstract its functionality as much as possible to allow its reusability in other areas.

The implementation of the plugin is based on the React Unity WebGL library, which allows embedding Unity compilations exported to WebGL format in any application developed on the React framework. Likewise, it allows establishing communications between the Unity model and the React application in both directions. The library provides a UnityContext object that is initialized with the four files that constitute the WebGL format. To access these files, the files must be placed in the Grafana public folder. To display the compilation, it is only necessary to reference this object in a JSX element that is also provided with the library. For the Unity model to work correctly, it is necessary to add in one of its scripts the disabling of the capture of all keyboard inputs, since otherwise it may interfere with the operation of the rest of the JavaScript elements that compose the application.

For receiving information in the Unity model coming from the React application, a function must first be defined in each of the Unity objects that wish to receive such information. This function will receive the data sent by the application as a parameter, typically as text or number. On the application side, the send method of the UnityContext object previously created must be used, indicating the name of the Unity object we are referring to, the name of the function mentioned above and, if any, the data to be sent. In our case, we want to send to the corresponding Unity objects the data obtained as a result of the query that the user has entered the Grafana panel. This information is contained in the data variable that Grafana provides as a parameter of the component. This variable contains a list of series, which will depend on how the data are grouped.. There shall be one series for each group in the query result. For example, if you group the data first by thingId and then by feature, and you have two twins with the same two features, then you will create 4 groups and therefore 4 series. Each series will contain a set of fields, and each field will have a list with its values. The values of the last grouping column will be considered as independent fields, while the values of the other grouping columns will be shown as labels of those fields. Thus, in the example above, each of the series would have a time field and a field with the name of the feature by which it is grouped, both labelled with the thingId to which they correspond. Taking all this into account, the plugin panel will extract the relevant data from this Grafana variable, analyse it and build the message for the device that is being filtered. Once sent, it will be received as a parameter in the corresponding Unity object function, which will act accordingly. This sending is done after each modification of the data variable, that is, after each reception of data in Grafana.

To send data from the Unity model to the React application, it is necessary to establish an event beforehand. In a specific folder inside the Unity model, a JSLib file has to be added where the name of the event and the type of parameters to be sent have to be defined. This event can be imported and used within Unity scripts that require sending information to the React application. To receive the data in the panel plugin, the corresponding method of the UnityContext object must be called, indicating the name of the event to wait for and the function to execute after receiving it. This function must have the parameters specified in the event. In our case, a script has been defined in Unity that, when a Unity object is clicked on, sends an event with its identifier. The identifiers of these objects must correspond to those used for the construction of the twin in Eclipse Ditto. This identifier will be set as the value of a variable of the Grafana dashboard where the panel is located. Grafana allows the definition of several types of variables that can be included in the queries to make the dashboards more dynamic and interactive. In this way, the rest of the dashboard panels will be able to display information depending on the Unity object that is clicked in the 3D model.

All variable elements of the panel, such as function name, event name, WebGL compilation location and dashboard variable name, have been defined as panel options to allow the user to easily modify these values.

As with the application plugin developed in a previous section, the developed code must be added to the Grafana plugins folder to be used. Once activated, it can be included in any Grafana dashboard, where it can be configured and adapted to the user's preferences.

\section{Use case: Virtual analyser in Petrochemical industry}
The digital twin platform here presented will be validated through an Petrochemical Industry 4.0 use case. The objective of the use case is to define a virtual analyzer, that is able to predict the freezing point of one of CEPSA end products (lubricant) based on the operating conditions and the properties of the feedstock. This process is carried out in the San Roque (Spain) Energy Park of CEPSA, one of the largest refinery in Spain. The freezing point is an important parameter that, due to its characteristics, must be measured in laboratory. Based on the monitoring of different operation conditions, such as filters, the aim is to predict in Kafka-ML the state of the freezing point in real time for better control of the process. This continuous prediction, together with the status of the monitored sensors, will be modelled in a digital twin within our framework. Digital twins have also been studied before in the Petrochemical Industry. For instance, in \cite{min2019machine}, a digital twin for production control purposes of a catalytic cracking unit in the Petrochemical Industry is proposed. In this work, we go further by considering the modelling, prediction, and 3D visualization for a process in this industry.

The company has provided us with real-time (through the MQTT protocol) and historical data (to train ML models) from the necessary sensors, which will be considered twins in their own right and together will constitute the main twin being sought. From them, the different digital twin types have been identified by grouping the sensors that are identical in operation and description, and a Ditto Thing scheme has been defined for each type and twin. In this case, all the sensors receive a single value along with the time it was taken, and, likewise, the format in which the data from each sensor is received is identical. In order to facilitate mapping and data consultation, the feature section is the same for all twins and types. Using the Digital Twins plugin for Grafana, connected to a Ditto-Extended-API service, the different twins have been created. All elements have been grouped in the same Eclipse Ditto namespace (a logical way to group information and digital twins). As an example, the Ditto Thing scheme for one of the sensors is shown in Figure \ref{fig:schemacepsa}.

\begin{figure}
\begin{lstlisting}[language=json]
{
  "thingId": "cepsa:LSRC3002.PF",
  "policyId": "cepsa:basic_policy",
  "attributes" : {
    "name": "LSRC3002.PF",
    "description" : "Unit load",
    "units" : "m3/d"
  },
  "features": {
    "last_measured": { 
      "properties": { 
        "value": null,
        "time": null
      } 
    }
  }
}
\end{lstlisting}
\caption{Example of the Eclipse Ditto schematic for one of the sensors in the use case.}
\label{fig:schemacepsa}
\end{figure}

For sending real-time data, a tenant was created in Eclipse Hono for the MQTT connection 
within which credentialed devices have been added for each of the available sensors. The name of each device coincides with the thingId of its corresponding twin, thus facilitating the connection with Eclipse Ditto. 
Moreover, in this same connection, a JavaScript mapper has been applied to convert the messages received into the Ditto Protocol format.

At this point, the twin should be receiving the data correctly, and the status of the twins with respect to time will be stored in InfluxDB, so Grafana dashboards can be created according to requirements.

The freezing point predictive model that has been developed as the target of the use case has been deployed in Kafka-ML. For its data input, a simple script has been written that periodically makes a call to Eclipse Ditto to collect the current state of the twins that represent the sensors required by the model. The output data of this model is collected by the Kafka-ML-to-Eclipse-Ditto service in order to update the freezing point feature contained in the digital twin.

For the 3D representation of the twin's state, a model was created in Unity that contains enough elements to represent each of the sensors that are part of the machine. In this model, each element has been renamed with the ID of the sensor it represents, and necessary code has been implemented for the movement of the camera and the selection of elements by clicking on them, as well as the script necessary for the model's correct working in Grafana. Its WebGL export has been added to the Grafana public folder, and the Unity panel has been included in one of the boards.

\begin{figure*}[ht] 
  \includegraphics[width=\linewidth]{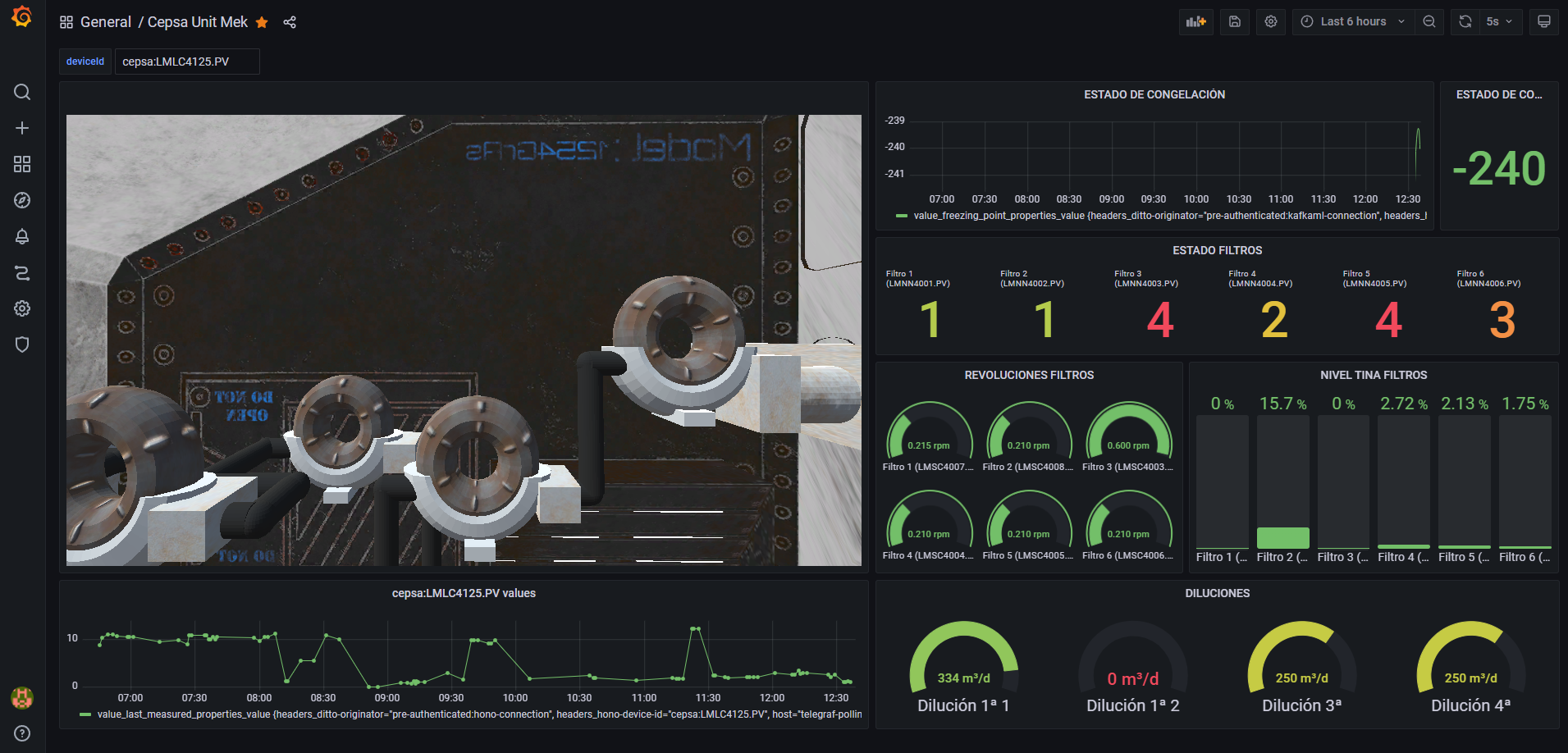}
  \caption{3D visualization of the digital twin for the use case}
  \label{fig:finalresult}
\end{figure*}

Figure \ref{fig:finalresult} shows the final result of the digital twin developed 3D representation for the described use case. As a result, an easily adaptable and extendable twin of the industry process has been obtained, with an eye-pleasing representation of its real-time status, using different types of graphics as well as a 3D model that, on receiving the data from the sensors, provides the possibility of displaying data of interest on the machine, such as its real movement, and allowing any type of interaction with the user. Likewise, the platform allows easy querying of the current state of the twin, via the Eclipse Ditto API, and its state over time, using any of the query options provided by InfluxDB. Furthermore, in terms of predictive concerns, machine learning models are easily integrated with the twins, with the resulting value being considered as any other feature of the twin. 

\section{Evaluation}
\label{sec:evaluation}

Next, a series of tests will be carried out to validate the performance and scalability through a latency analysis, as well as the availability of the platform using the construction outlined in the previous section and taking into account the different flows of which the architecture is composed.

Test 1 will probe the latency and throughput of the digital twin's core functionalities flow with respect to the number of connected sensors and the number of clients/connections used. Meanwhile, Test 2 will calculate the same properties but will focus on the predictive flow of the platform considering the number of simultaneous clients. Finally, Test 3 will check the fault tolerance of the platform by obtaining the average recovery time of the services and whether the failure of those services results in data loss.

\subsection{Experimental setup}
\label{sec:setup}
\textbf{Hardware configuration}. All the experiments were performed on a five-node Kubernetes cluster in our private cloud infrastructure in VMware vCloud. Each node has 4 virtual CPUs in 2 sockets and 16GB of RAM. The client that sent the information and from where the results were measured was a PC with 64GB of RAM, 1 CPU, and 10 cores. 

\textbf{Software configuration}. Each one of the five nodes runs Kubernetes v1.19.3 and Docker 19.03.13 on top of Ubuntu 16.04.7 LTS. A Kubernetes master was deployed in one node, whereas the remaining four are Kubernetes workers. The PC with the client runs Ubuntu server.

\subsection{Test 1 - Essential functionality flow}

The test will evaluate the latency and throughput from the moment data are sent to Eclipse Hono via MQTT to the moment they are stored in InfluxDB within the dataflow used in the Petrochemical Industry use case described. Here we have two test cases, different number of sensors receiving data simultaneously, and different numbers of clients/connections sending data simultaneously. The size of the data sent has not been taken into account, because all the messages share the same format, so their size hardly varies.

\subsubsection{Related to number of sensors}

\begin{figure*}[h]
  \centering 
  \subfloat[][Latency]{\includegraphics[width=0.4\linewidth]{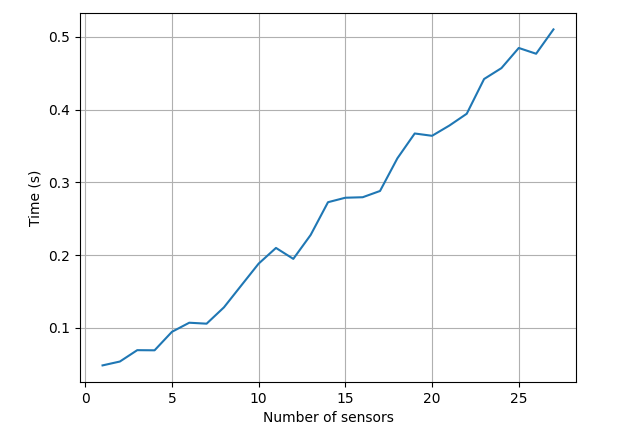}} 
  \qquad 
  \subfloat[][Throughput]{\includegraphics[width=0.4\linewidth]{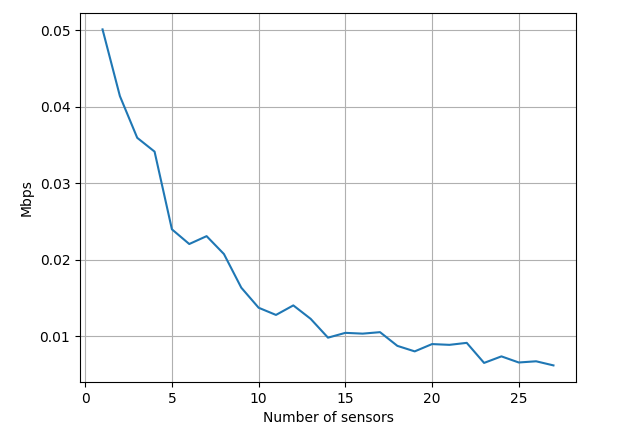}} 
  \caption{Latency and throughput of test 1 for different number of sensors}
  \label{fig:results_test1_sensors}
\end{figure*}

For this test, historical data have been sent in the format specified by the company, creating an MQTT connection for each sensor and increasing the number of sensors to which data are sent simultaneously using threads. The time at which each input is stored in InfluxDB is then queried, and times are compared to obtain latency and throughput. The result of each test by sensor number is the average of ten repetitions of the test. Since the company provided us with data from 27 of their sensors, this is the maximum number of sensors to be called simultaneously in this test. Figure \ref{fig:results_test1_sensors} shows the results obtained.

It is clear that, as the number of affected sensors increases, latency increases almost linearly, and throughput decreases exponentially. This result is in accordance with normality, and it should also be taken into account that in a real scenario not all the sensors of the platform will receive the data simultaneously or with the same frequency, so we can determine that the platform performs well as the number of sensors (here twins) affected increases.

\subsubsection{Related to number of clients}

\begin{figure*}[h]
  \centering 
  \subfloat[][Latency]{\includegraphics[width=0.4\linewidth]{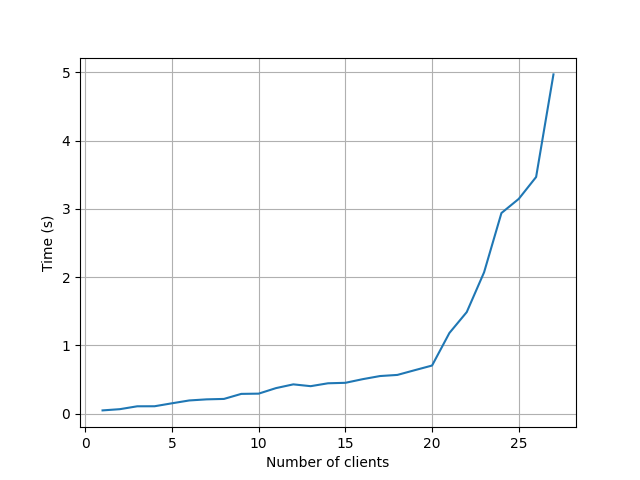}} 
  \qquad 
  \subfloat[][Throughput]{\includegraphics[width=0.4\linewidth]{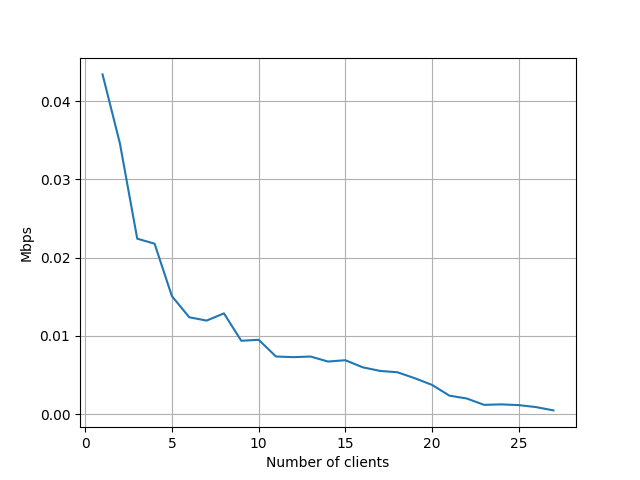}} 
  \caption{Latency and throughput of test 1 for different number of clients}
  \label{fig:results_test1_clients}
\end{figure*}

This case is similar to the previous one, with the difference that only data updates will be sent to a single sensor by a different number of simulated clients using threads. The values sent in each message are always unique, as each client increments by 0.01 a global value, starting at 0. Likewise, the result of each test per number of clients is the average of ten repetitions of this test. Figure \ref{fig:results_test1_clients} shows the results, which have also been limited to 27 clients to facilitate the understanding of the graph and the comparison with the one explained above.

As can be seen, the latency exceeds one second of delay after 20 simultaneous clients. The throughput, on the other hand, maintains a fair decrease. These results do not represent a problem, since the most common stage is that a twin or device receives data from a single data source, with the exception of simulated or predicted features, where the number of clients could increase usually by one or two. This behaviour is also normal as sending all messages to a single twin can overload it. Therefore, we can conclude that the system reacts correctly to a coherent increase in the number of clients.

\subsection{Test 2 - Machine learning prediction flow}

\begin{figure*}[h]
  \centering 
  \subfloat[][Latency]{\includegraphics[width=0.4\linewidth]{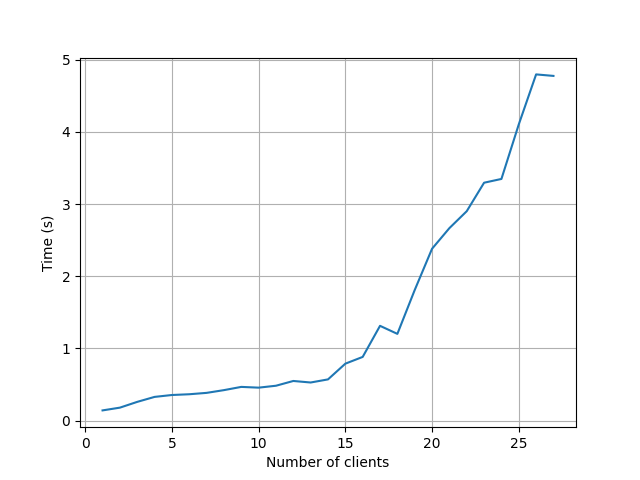}} 
  \qquad 
  \subfloat[][Throughput]{\includegraphics[width=0.4\linewidth]{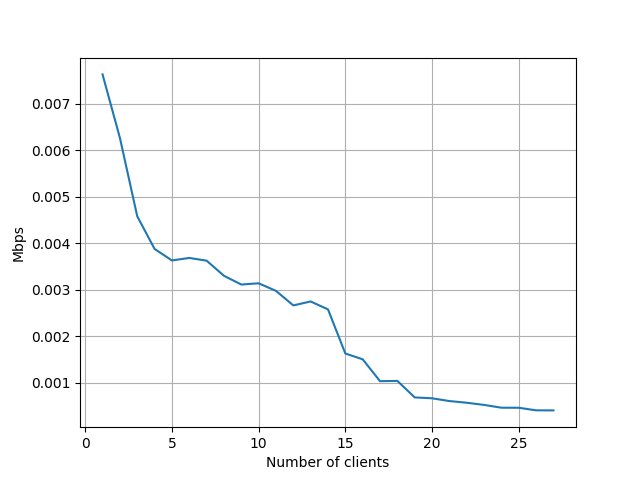}} 
  \caption{Latency and throughput of test 2}
  \label{fig:results_test2}
\end{figure*}

In this test the latency and throughput will be calculated for the machine learning integration part of the architecture. Since there is only one prediction model running in the use case for the freezing point prediction, the test will always affect the same Eclipse Ditto twin, and what we will vary is the number of clients sending input data to the model. To relate each client to the outcome of the model, we have used data inputs with known results, avoiding their repetition during each of the tests. After the execution of the test with a certain number of clients, InfluxDB is consulted for the time in which each data has been stored, which will allow later comparison. The result of each test is the average of 10 executions. Figure \ref{fig:results_test2} shows the results achieved in this test.

The results are very similar to those shown in the first test with respect to the number of clients. Latency grows until it exceeds one second delay with 17 simultaneous clients. Similarly, throughput decreases substantially. These are acceptable results compared to the real-time data flow, as the difference between them is minor, and it must be taken into account that a per-client prediction is made during the process. Moreover, in this case, as in the previous one, it is usual for a single client to initiate the flow, so the limitation of simultaneous clients would not be a problem. Therefore, we can determine that the platform has a very good response to the increase in the number of simultaneous clients during the prediction flow through machine learning.

\subsection{Test 3 - Error tolerance}

\begin{figure}[h]
  \includegraphics[width=\linewidth]{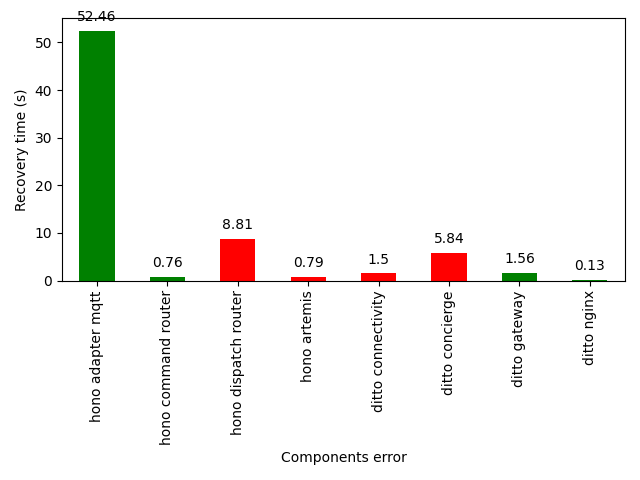}
  \caption{Recovery time for each service}
  \label{fig:test4}
\end{figure}

This last test is based on checking the platform's tolerance to errors. To do this, a script has been created that sends a message to Eclipse Hono via MQTT every half a second. While this is running, the pod corresponding to the service to be tested is manually deleted in Kubernetes. If a message could not be sent, it will be resent as many times as necessary, although maintaining the initial time of the first sending. When the end of the test is indicated, the data received will be extracted from InfluxDB for comparison. The maximum time difference of the test shall be considered as the recovery time for that pod. Each test result is the average of 5 test runs. In Figure \ref{fig:test4}, we can see the results. The selected pods follow the sequence from sending a data to Eclipse Hono until the moment the corresponding twin is updated in Eclipse Ditto and are shown in the graph in that order. The combination of two or more pods has not been included as the result coincides with the maximum recovery time between them.

First of all, it should be noted that these results are highly dependent on the capabilities of the cluster where the services are deployed. It can be seen that the recovery time of the services tends to be very low, which indicates very good results. The exception is the Eclipse Hono MQTT adapter, which has a higher recovery time, probably owing to its dependence on the creation, configuration, and connection of an MQTT broker. On the other hand, by looking at the raw data obtained in the test, we can determine which services involve a loss of data during recovery. Those that do involve some data loss are marked in red in Figure \ref{fig:test4}, while those that do not are shown in green. This is not a big problem, because of the short recovery time of the services, although its resolution will be studied for future work. Based on the above, we can conclude that the platform has got a good behaviour with respect to errors, as it is capable of recovering from them without causing a great loss of data or requiring human intervention.

\section{Conclusions and future work}

Digital twins are emerging as a valuable resource in order to have a better understanding of and anticipate possible situations that assets may face in the physical world. Although a multitude of platforms have been defined in the literature for the development of digital twins, they are mainly focused on specific vertical contexts, and a significant evolution is needed to achieve effective digital twins. In this paper, we propose an open-source platform for the development of effective digital twins that can be adapted to multiple contexts. We refer to digital twins that can be seamlessly integrated with predictive models through the open platform Kafka-ML, interactive digital twins provided thanks to Unity's support for 3D representations, and a unified interface for real-time visualization and management of IoT monitoring data for simple and complex models. To demonstrate the feasibility of the platform, we have defined a digital twin of an industrial process in the Petrochemical Industry. This process monitors and predicts the freezing point in lubricant generation. Through the digital twin, plant users can visualize the status of the plant and its components (e.g. filters) directly on the 3D representation, as well as the freezing point prediction in real time.

As future work for the platform, we envisage supporting the FMI (Functional Mock-up Interface) standard in order to be able to simulate complex processes that follow this standard and integrate them with the 3D representations of the platform. Furthermore, we intend to generate hybrid twins that can take advantage of the low latency opportunities offered by edge/fog systems and demonstrate the viability of the platform in other contexts (we are currently working in the agricultural sector).     

\label{sec:conclusions-future}

 \bibliographystyle{elsarticle-num} 
 \bibliography{main}

\begin{thebibliography}{10}
\expandafter\ifx\csname url\endcsname\relax
  \def\url#1{\texttt{#1}}\fi
\expandafter\ifx\csname urlprefix\endcsname\relax\def\urlprefix{URL }\fi
\expandafter\ifx\csname href\endcsname\relax
  \def\href#1#2{#2} \def\path#1{#1}\fi

\bibitem{diaz2016state}
M.~D{\'\i}az, C.~Mart{\'\i}n, B.~Rubio, State-of-the-art, challenges, and open
  issues in the integration of internet of things and cloud computing, Journal
  of Network and Computer applications 67 (2016) 99--117.

\bibitem{de2020toward}
D.~De~Silva, S.~Sierla, D.~Alahakoon, E.~Osipov, X.~Yu, V.~Vyatkin, Toward
  intelligent industrial informatics: A review of current developments and
  future directions of artificial intelligence in industrial applications, IEEE
  Industrial Electronics Magazine 14~(2) (2020) 57--72.

\bibitem{tao2018digital}
F.~Tao, H.~Zhang, A.~Liu, A.~Y. Nee, Digital twin in industry:
  State-of-the-art, IEEE Transactions on Industrial Informatics 15~(4) (2018)
  2405--2415.

\bibitem{rasheed2020digital}
A.~Rasheed, O.~San, T.~Kvamsdal, Digital twin: Values, challenges and enablers
  from a modeling perspective, Ieee Access 8 (2020) 21980--22012.

\bibitem{marquez2020designing}
A.~C. M{\'a}rquez, A.~de~la Fuente~Carmona, J.~A. Marcos, J.~Navarro, Designing
  cbm plans, based on predictive analytics and big data tools, for train wheel
  bearings, Computers in Industry 122 (2020) 103292.

\bibitem{nazarenko2020role}
A.~A. Nazarenko, L.~M. Camarinha-Matos, The role of digital twins in
  collaborative cyber-physical systems, in: Doctoral Conference on Computing,
  Electrical and Industrial Systems, Springer, 2020, pp. 191--205.

\bibitem{conde2021modeling}
J.~Conde, A.~Munoz-Arcentales, A.~Alonso, S.~Lopez-Pernas, J.~Salvachua,
  Modeling digital twin data and architecture: A building guide with fiware as
  enabling technology, IEEE Internet Computing (2021).

\bibitem{zheng2019application}
Y.~Zheng, S.~Yang, H.~Cheng, An application framework of digital twin and its
  case study, Journal of Ambient Intelligence and Humanized Computing 10~(3)
  (2019) 1141--1153.

\bibitem{cheng2020dt}
J.~Cheng, H.~Zhang, F.~Tao, C.-F. Juang, Dt-ii: Digital twin enhanced
  industrial internet reference framework towards smart manufacturing, Robotics
  and Computer-Integrated Manufacturing 62 (2020) 101881.

\bibitem{mo2021terra}
Y.~Mo, S.~Ma, H.~Gong, Z.~Chen, J.~Zhang, D.~Tao, Terra: A smart and sensible
  digital twin framework for robust robot deployment in challenging
  environments, IEEE Internet of Things Journal 8~(18) (2021) 14039--14050.

\bibitem{dang2021cloud}
H.~V. Dang, M.~Tatipamula, H.~X. Nguyen, Cloud-based digital twinning for
  structural health monitoring using deep learning, IEEE Transactions on
  Industrial Informatics (2021).

\bibitem{kamath2020industrial}
V.~Kamath, J.~Morgan, M.~I. Ali, Industrial iot and digital twins for a smart
  factory: An open source toolkit for application design and benchmarking, in:
  2020 Global Internet of Things Summit (GIoTS), June 3, Online, IEEE, 2020,
  pp. 1--6.

\bibitem{shah2021construction}
K.~Shah, T.~Prabhakar, C.~Sarweshkumar, S.~Abhishek, et~al., Construction of a
  digital twin framework using free and open-source software programs, IEEE
  Internet Computing (2021).

\bibitem{rolle2021modular}
R.~P. Rolle, V.~d.~O. Martucci, E.~P. Godoy, Modular framework for digital
  twins: Development and performance analysis, Journal of Control, Automation
  and Electrical Systems 32~(6) (2021) 1485--1497.

\bibitem{khan2020toward}
A.~Khan, F.~Shahid, C.~Maple, A.~Ahmad, G.~Jeon, Toward smart manufacturing
  using spiral digital twin framework and twinchain, IEEE Transactions on
  Industrial Informatics 18~(2) (2020) 1359--1366.

\bibitem{martin2022kafka}
C.~Mart{\'\i}n, P.~Langendoerfer, P.~S. Zarrin, M.~D{\'\i}az, B.~Rubio,
  Kafka-ml: connecting the data stream with ml/ai frameworks, Future Generation
  Computer Systems 126 (2022) 15--33.

\bibitem{min2019machine}
Q.~Min, Y.~Lu, Z.~Liu, C.~Su, B.~Wang, Machine learning based digital twin
  framework for production optimization in petrochemical industry,
  International Journal of Information Management 49 (2019) 502--519.

\end{thebibliography}

\par\noindent 
\parbox[t]{\linewidth}{
\noindent\parpic{\includegraphics[height=1.2in,width=0.8in,clip,keepaspectratio]{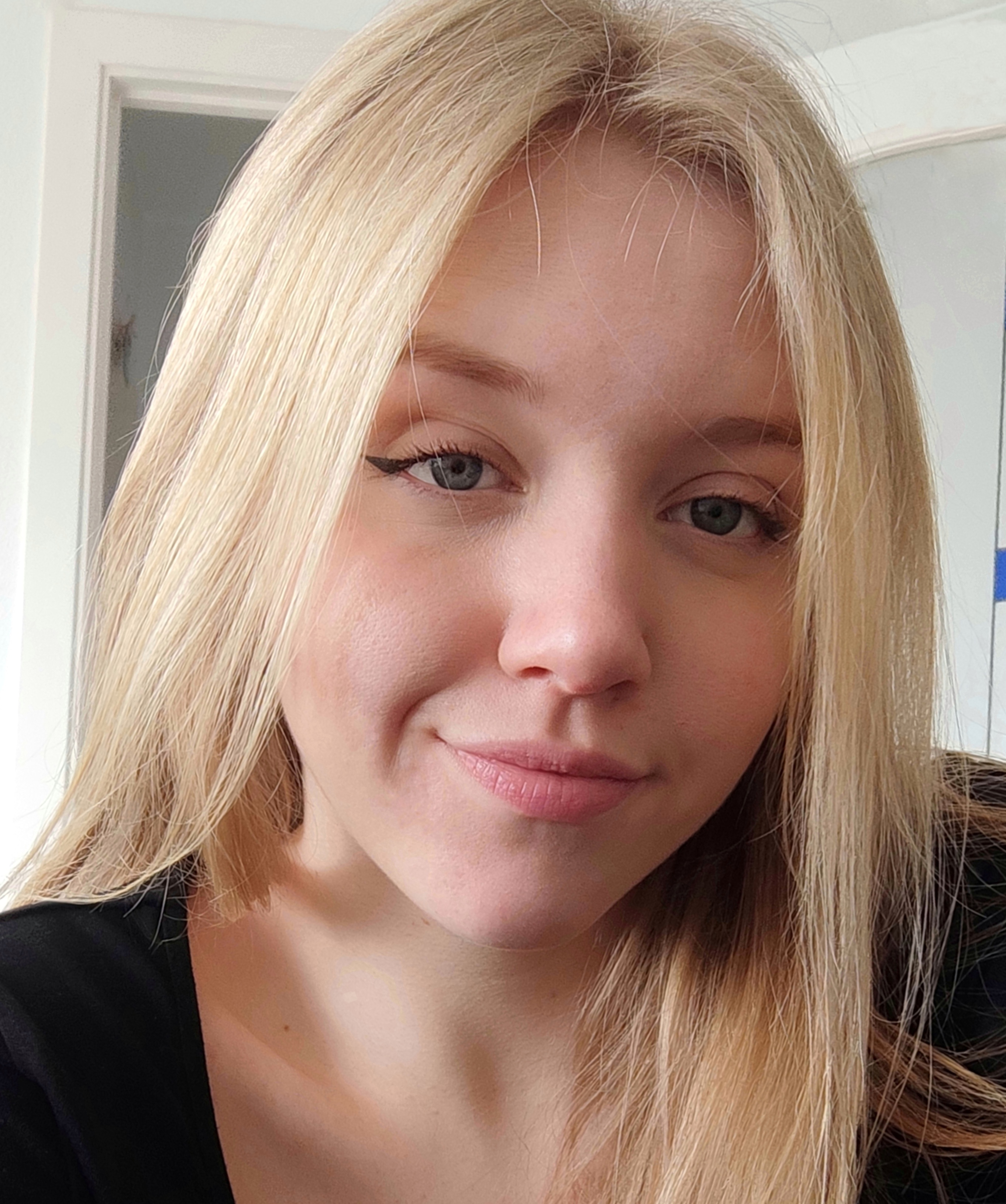}}
\noindent {\bf Julia Robles}\
graduated in Software Engineering from the University of Málaga, Spain, in 2021. Since then, she is part of the ERTIS Research Group at the University of Málaga as a research assistant and is a member of the ITIS Software Institute at the University of Málaga. His research interests are mainly in the area of Digital Twins, Internet of Things, and Artificial Intelligence.}
\vspace{4\baselineskip}

\par\noindent 
\parbox[t]{\linewidth}{
\noindent\parpic{\includegraphics[height=1.2in,width=0.8in,clip,keepaspectratio]{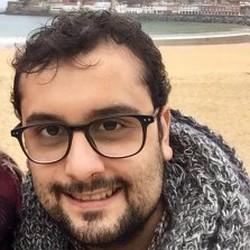}}
\noindent {\bf Cristian Martín}\
received an MS in Computer Engineering, an MS in Software Engineering and Artificial Intelligence, and a PhD in Computer Science from the University of Málaga, Spain, in 2014, 2015, and 2018 respectively. Currently, he is a Postdoctoral researcher at the University of Málaga. Previously, he has worked as a software engineer in various tech companies with RFID technology and software development. He is also a member of the ITIS Software Institute of the University of Málaga. His research interests focus on the integration of the Internet of Things with Cloud/Fog/Edge Computing, Machine Learning, Structural Health Monitoring, and IoT Reliability.}
\vspace{4\baselineskip}

\par\noindent 
\parbox[t]{\linewidth}{
\noindent\parpic{\includegraphics[height=1.2in,width=0.8in,clip,keepaspectratio]{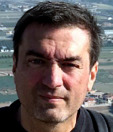}}
\noindent {\bf Manuel Díaz}\
is Full Professor in the Computer Science Department at the University of Málaga and Head of the ERTIS research group and member of the ITIS software Institute. His research interests are in distributed and real-time systems, Internet of Things, and P2P, especially in the context of middleware platforms and critical systems. In the last years, his main area of work have been WSN and monitoring systems, especially energy monitoring (FP7 e-balance project), water infrastructure monitoring (FP7 SAID project), and energy efficient buildings (FP7 SEEDS). He has collaborated in many technology transfer projects with different companies such as Tecnatom, Telefónica, Indra, or Abengoa. He was the coordinator of the FP6 SMEPP project and main researcher for UMA in the WSAN4CIP (ICT FP7). He is also co-founder of the spin-off Softcrits and head of its R\&D department.}
\vspace{4\baselineskip}





\end{document}